\documentclass[10pt,conference]{IEEEtran}
\usepackage{epsfig,setspace,amsmath,epsf,amssymb,bm,theorem,cite,graphicx, epstopdf,algorithm,float,color,mathtools, authblk,physics}
\usepackage[table,xcdraw]{xcolor}
\usepackage{subcaption}
\usepackage{algorithm,algorithmic}
\usepackage{dsfont}
\usepackage[short,c2]{optidef}

\newtheorem{theorem}{Theorem}

\newtheorem{lemma}{Lemma}

\newtheorem{proposition}{Proposition}

\newcommand{\e}{{\mathbb{E}}}

\IEEEoverridecommandlockouts
\allowdisplaybreaks

\begin{document}

\title{Preemption Revisited: Multi-Threshold Preemption Policies for AoI Minimization}

\author[1]{Sahan Liyanaarachchi}
\author[1]{Sennur Ulukus}
\author[2]{Nail Akar}

\affil[1]{\normalsize University of Maryland, College Park, MD, USA}
\affil[2]{\normalsize Bilkent University, Ankara, T\"{u}rkiye}

\maketitle
\begin{abstract}
    The study of optimal preemption policies for status update systems has been a recurring topic in the age of information (AoI) literature, where threshold-based structures have been shown to be optimal under a generate-at-will update generation model under certain assumptions. In this work, we study the effectiveness of threshold-based policies for a system with random update arrivals. In this regard, we introduce an analytical framework for evaluating the AoI of multi-threshold preemption policies and present interesting characteristics of the structure of the optimal preemption policy. We show the effectiveness of these threshold-based policies over the traditional probabilistic preemption policies and single-threshold policies, where we observe that significant gains in terms of AoI can be obtained by utilizing both the age of the packet and the age of the system when designing these preemption policies.
\end{abstract}

\section{Introduction}
In this new era of modern computing, where the sheer volume of data processed in a single time unit is extreme, one fundamental problem lingers\textemdash upon arrival of a new job request, should we forgo our invested efforts on the current job and focus our efforts to this new job, i.e., preempt the job in service? This is a fundamental problem rooted in queuing theory and have been widely studied in the realm of status update systems, where service time of the queue models the delay experienced by the updates\cite{status_update_queues}. In such systems, the decisions on whether to preempt or not are made to maximize the freshness of the updates, which is often quantified through the age of information (AoI) metric \cite{yates2020age}. 

This problem has been studied throughout the literature in multiple occasions. In \cite{subhankar_preempt}, the authors aim to find the structure of the optimal preemption policy in the discrete-time domain under a generate-at-will model. They show that, under certain conditions, the optimal preemption policy has a double-threshold structure. The work in \cite{elif_heavytail} extends this study  to the continuous-time domain with an impulse control framework. They show that threshold structures emerge naturally under a generate-at-will model.
Under a random arrival update generation model, the work in \cite{prob_preempt} highlights the importance of probabilistic preemption policies for certain delay distributions. Reference \cite{link_capacity_preempt} studies the problem of preemption under a random arrival model with the assumption that the intended service time of the new arrival is known before hand. Under this assumption, they show that threshold structures emerge in the optimal preemption policies. The work in \cite{delayed_premption} looks into preemption policies based on a single fixed threshold under Poison arrivals. On a similar note, under a generate-at-will model, reference \cite{timely_computing_preempt} focuses on finding optimal sampling policies by restricting preemption to a single-threshold policy. 

These works highlight the importance of threshold-based preemption policies. However, most of these works are restricted to single- or double-threshold policies for simplicity or focus on distributions where these policies are optimal. However, for general service time distributions, the optimal preemption policies are more complex. In fact, under certain service time distributions, the optimal preemption policies may be a mixture of multiple threshold policies depending on both the age of the system and the age of the packet in transmission. In this work, we restrict ourselves to a finite mixture of threshold policies. Under these policies, we give an analytical framework based on absorbing Markov chains (AMCs) to find the average AoI. We also give certain structural characteristics of these multi-threshold preemption policies. Using this framework, we show that these multi-threshold preemption policies outperform conventional probabilistic and single-threshold preemption policies. 

\section{System model}
Consider a status update system operating in the discrete-time domain, where new packets are generated at the beginning of a time-slot with probability $q$. We assume that the system is bufferless and hence any newly generated packet will be discarded if not subjected to the channel for transmission. The packets are transmitted across a random delay channel, where the discrete random variable $Y$ represents the experienced delay. We assume that all channel transmissions occur at the end of a given time-slot. Let $\Delta_n=n-U_n$ denote the age of the system, i.e., the AoI at the receiver, where $U_n$ is the time-slot in which the most recently received packet was generated. Let $\delta_n=n-G_n$ denote the age of the packet (AoP) that is currently in transmission, where $G_n$ denotes its generation time. If there is no packet in transmission, we say that the system is idle and AoP is not defined in this instance.

As a control policy, we enforce the following preemption policy for the above system. Let $\{\Gamma^{(0)},\Gamma^{(1)},\dots,\Gamma^{(N)}\}$ be a set of thresholds with $\Gamma^{(0)}=0$, $\Gamma^{(N)}=\infty$ and $\Gamma^{(i-1)}<\Gamma^{(i)}$ for $1\leq i\leq N$. If $\Gamma^{(i-1)}\leq\Delta_n< \Gamma^{(i)}$ and $\delta_n=k$, we initiate a preemptive transmission with probability $p^{(i)}_k$ upon the generation of a new packet. If the system is idle, then a newly generated packet will be subjected to transmission with probability $p^{(i)}_0$ if $\Gamma^{(i-1)}\leq\Delta_n< \Gamma^{(i)}$. We assume that any packet whose AoP goes beyond $M$ is too stale and conveys no useful information to the receiver. Hence, if $\delta_n>M$, that particular packet is discarded from the system and the system goes to the idle state. Now, we will find the average AoI of the system under the above class of preemption policies.

\section{AoI Analysis}
To model the AoI process, we consider one AoI cycle when in steady state. We consider a cycle which starts and ends with the successful reception of a packet under the above preemption policy. Let $\Delta_n=\gamma$ at the beginning of this cycle. Since we discard packets whose age is greater than $M$, we have at most $M$ possibilities for $\gamma$. For a given $\gamma$, we first find the average AoI of the cycle by constructing an absorbing Markov chain (AMC) which begins when the system is idle and gets absorbed when a packet is successfully delivered to the receiver. Let $\tau_\gamma$ denote the absorption time. Then, the cycle duration for this AoI cycle is $\tau_\gamma$ and the AoI increases from $\gamma$ to $\gamma+\tau_\gamma-1$ in this particular cycle. Therefore, if $\pi_i=\mathds{P}(\gamma=i)$, then the average AoI denoted by $\bar\Delta$ can be found as follows,
\begin{align}
    \bar\Delta=\frac{\sum_{i=1}^{M}\pi_i\left(2i\e[\tau_i]+\e[\tau_i^2]\right)}{2\sum_{i=1}^{M}\pi_i\e[\tau_i]}-\frac{1}{2}.\label{eqn:avg_age}
\end{align}
Thus, to find $\bar\Delta$, we only need to find the pmf of $\gamma$ and the first two moments of $\tau_\gamma$.

The AMC that models $\tau_\gamma$  is composed of $M+1$ transient states one for each packet age and one for the idle state of the system. It will have $M$ absorbing states depending on the age of the packet at the time of absorption. Since our preemption policy varies with $\Delta_n$, so does the probability transition matrix of our AMC. Let $\bm{P}_i$ denote the probability transition matrix of our AMC when $\Gamma^{(i-1)}\leq\Delta_n< \Gamma^{(i)}$. Now, $\bm{P}_i$ can be represented as follows,
\begin{align}
    \bm{P}_i=\begin{bmatrix}
        \bm{S}_i &\bm{A}_i\\
        \bm{0}&\bm{I}_M
    \end{bmatrix},
\end{align}
where $\bm{S}_i$ is an $(M+1)\times (M+1)$ matrix corresponding to the transitions between transient states, $\bm{A}_i$ is an $(M+1)\times M$ matrix corresponding to the transitions from transient states to absorbing states, $\bm{I}_M$ is an identity matrix of dimension $M$, and $\bm{0}$ is a matrix of all zeros of appropriate dimension. 

Let the states of the above transition matrix be ordered as $\{1_S,2_S,\dots,M_S,idle,1_A,2_A,\dots,M_A\}$, where $i_S$ a is transient state with $i$ denoting the age of the packet under transmission, $i_A$ is an absorbing state with $i$ denoting the age of the packet when the packet is successfully received, and $idle$ is a transient state which represents that there is no packet under transmission. Let $y_n=\mathds{P}(Y=n+1|Y>n)$ be the conditional probability that the current transmission will be successful in the next time-slot, given $n$ time units have elapsed. For brevity, we will drop the dependence on $i$ and define  $p^{(i)}_kq=q_k$  and the operator $\bar x=1-x$. Then, the matrices $\bm{S}_i$ and $\bm{A}_i$ can be expressed as follows,
\begin{align}
   \bm{S}_i&=\begin{bmatrix}
       q_1\bar y_0&\bar q_1\bar y_1&0&\cdots&0&0\\
       q_2\bar y_0&0&\bar q_2\bar y_2&\cdots&0&0\\
       \vdots&\vdots&\vdots &\ddots&\vdots&\vdots\\
       q_{M-1}\bar y_0&0&0&\cdots&\bar q_{M-1}\bar y_{M-1}&0\\
       q_M\bar y_0&0&0&\cdots&0&\bar q_M\\
       q_0\bar y_0&0&0&\cdots&0&\bar q_0\\
   \end{bmatrix}, \\
   \bm{A}_i&=\begin{bmatrix}
       q_1y_0&\bar q_1 y_1&0&\dots &0\\
       q_2y_0&0&\bar q_2y_2&\dots&0\\
       \vdots&\vdots&\vdots&\ddots&\vdots\\
       q_{M-1}y_0&0&0&\cdots&\bar q_{M-1}y_{M-1}\\
       q_My_0&0&0&\cdots&0\\
       q_0y_0&0&0&\cdots&0
   \end{bmatrix}.
\end{align}
Note, since $\bm P_i$ is a stochastic matrix, we have $\bm{S}_i\bm{1}+\bm{A}_i\bm{1}=\bm{1}$, where $\bm{1}$ represents a  column vector of all ones of appropriate dimension. Now, the distribution of $\tau_\gamma$ is given in Proposition \ref{prop:dist_tau_gamma} and its proof is presented in Appendix \ref{appen:prop_dist_tau_gamma}.

\begin{proposition}\label{prop:dist_tau_gamma}
    The distribution of $\tau_\gamma$ is given by,
    \begin{align}
        \mathds{P}(\tau_\gamma> n)=\bm\alpha_0\left(\prod_{k=0}^{i-1}\bm S_k^{\zeta_\gamma^{(k)}} \right)\bm S_i^{n-\Gamma_\gamma^{(i-1)}}\bm{1},\label{eqn:CDF}
    \end{align}
    for $\Gamma_\gamma^{(i-1)}<n\leq\Gamma_\gamma^{(i)}$, where $\Gamma_\gamma^{(i)}=(\Gamma^{(i)}-\gamma)^+$, $\zeta_\gamma^{(i)}=\Gamma_\gamma^{(i)}-\Gamma_\gamma^{(i-1)}$, $\bm{S}_0=\bm I_{M+1}$, $\zeta_\gamma^{(0)}=0$, and $\bm\alpha_0$ is a $M+1$ dimensional row vector of zeros with a one in the last entry.
\end{proposition}

Now, with $\mathds{P}(\tau_\gamma>n)$ at hand, we can evaluate the first and second moments of $\tau_\gamma$ using Lemma \ref{lem:moments} below.

\begin{lemma}\label{lem:moments}
    The first and second moments of $\tau_\gamma$ is given by,
    \begin{align}
        \e[\tau_\gamma]&= \sum_{k=0}^\infty \mathds{P}(\tau_\gamma>k),\label{eqn:1_moment}\\
        \e[\tau_\gamma^2]&= \e[\tau_\gamma]+2\sum_{k=1}^\infty k \mathds{P}(\tau_\gamma>k).\label{eqn:2_moment}
    \end{align}
\end{lemma}

Lemma \ref{lem:moments} follows from a standard result from probability theory. Now, substituting for $\mathds{P}(\tau_\gamma>k)$  and simplifying the expression in Lemma \ref{lem:moments} using matrix geometric series, yields the following set of equations, 
\begin{align}
    \e[\tau_\gamma]=&1+\bm\alpha_0\sum_{l=1}^{N-1}\left[\left(\prod_{k=0}^{l-1}\bm S_k^{\zeta_\gamma^{(k)}} \right)\sum_{k=1}^{\zeta_\gamma^{(l)}}\bm S_l^k\right]\bm1\nonumber\\
    &+\bm \alpha_0\left(\prod_{k=0}^{N-1}\bm S_k^{\zeta_\gamma^{(k)}} \right)\bm S_N(\bm I-\bm S_N)^{-1}\bm1,\\
    \e[\tau_\gamma^2]=&\e[\tau_\gamma]+2\bm{\alpha}_0\left(\prod_{k=0}^{N-1}\bm S_k^{\zeta_\gamma^{(k)}} \right)\bm S_N(\bm I-\bm S_N)^{-2}\bm1\nonumber\\
    &+2\Gamma^{(N-1)}_{\gamma}\bm{\alpha}_0\left(\prod_{k=0}^{N-1}\bm S_k^{\zeta_\gamma^{(k)}} \right)\bm S_N(\bm I-\bm S_N)^{-1}\bm1\nonumber\\
    &+2\bm\alpha_0\sum_{l=1}^{N-1}\left[\left(\prod_{k=0}^{l-1}\bm S_k^{\zeta_\gamma^{(k)}} \right)\sum_{k=1}^{\zeta_\gamma^{(l)}}(k+\Gamma^{(l-1)}_\gamma)\bm S_l^k\right]\bm1.
\end{align}

Next, in Proposition \ref{prop:pi}, we show that $\pi_i$s are the solution to the unique stationary distribution of a stochastic matrix. The proof of Proposition \ref{prop:pi} is given in Appendix \ref{appen:prop_pi}.

\begin{proposition}\label{prop:pi}
    Let $\bm \pi=\{\pi_1,\pi_2,\dots,\pi_M\}$ be a row vector. Then, $\bm \pi$ satisfies the following,
    \begin{align}
        \bm \pi=\bm \pi\bm B,
    \end{align}
    where $B$ is a stochastic matrix and its elements are,
    \begin{align}
        B_{ij}=&\bm\alpha_{0}\sum_{l=1}^{N-1}\left[\left(\prod_{k=0}^{l-1}\bm S_k^{\zeta_i^{(k)}} \right)\sum_{k=0}^{\zeta_i^{(l)}-1}\bm S_l^k\bm A_l\right]\bm \beta_j\nonumber\\
        &+\bm\alpha_{0}\left(\prod_{k=0}^{N-1}\bm S_k^{\zeta_i^{(k)}} \right)(\bm I- \bm S_N)^{-1}\bm A_N\bm \beta_j
    \end{align}
    with $\bm \beta_i$ being an $M$ dimensional column vector of all zeros except for a one at the $i$th position.
\end{proposition}

\section{Preemption Policies}\label{sec:prem_pol}
In this section, we identify interesting characteristics of an optimal preemption policy and formally introduce the low-complexity probabilistic and threshold policies considered in this work. 

\begin{theorem}\label{thrm:deter}
    If $N=\infty$, then there exists an optimal deterministic policy (i.e., $p^{(i)}_k\in\{0,1\}$).
\end{theorem}

\begin{theorem}\label{thrm:always_preempt}
    If $q=1$ and $y_n\leq y_0$ for $ 0<n<M$, then the always preempt policy is optimal. 
\end{theorem}

Theorem \ref{thrm:always_preempt} is identical to the result in \cite{subhankar_preempt} and may serve as an alternative proof that does not require an MDP formulation. The proofs of Theorem \ref{thrm:deter} and Theorem \ref{thrm:always_preempt} are given in the Appendix \ref{appen:thrm_deter} and Appendix \ref{appen:thrm_always_preempt}, respectively.

\subsection{Probabilistic Preemption (PP)}
In this policy, if the channel is idle, then any new generated packet will be transmitted with probability $1$ and if the channel is currently transmitting a packet when a new packet arrives, then this old packet will be preempted with probability $p^*$. The optimal $p^*$ is found using an exhaustive search.

\subsection{Packet Age-based Preemption (PAP)}
In this policy, we consider that the preemption policy only depends on AoP and hence $N=1$ in this setting. Moreover, to simplify the policy further, we consider that $p_0=1$, and $p_k=p^*_1$ for $1\leq\delta_n<\delta'$ and  $p_k=p^*_2$ for  $\delta'\leq\delta_n\leq M$. The optimal parameters for $p^*_1$ and $p^*_2$ and $\delta'$ are found through an exhaustive search. 

\subsection{Packet and System age-based Preemption (PSP)}
In this policy, we consider that the policy depends on both AoP and AoI. However, to simplify the policy, we only consider a single-threshold, i.e., $N=2$, and moreover, we restrict our preemption probabilities to be deterministic. In particular, if $0<\Delta_n<\Gamma$ then $p_k=p^*_1$ for $1\leq\delta_n<\delta_1'$ and  $p_k=p^*_2$ for  $\delta'_1\leq\delta_n\leq M$. If $\Gamma\leq\Delta_n<\infty$ then $p_k=p^*_3$ for $1\leq\delta_n<\delta_2'$ and  $p_k=p^*_4$ for  $\delta_2'\leq\delta_n\leq M$. Further, $p_i^*\in \{0,1\}$ for $i=1$ to $4$ and $p_0^1=p_o^2=1$. The optimal parameters are found using an exhaustive search, where we have confined the search space for $\Gamma$ to be less than $3M$.

Even though we use exhaustive search to find the optimal parameters for these policies, when the parameter space is large, since we have the closed-form expressions for $\bar\Delta$, one may resort to numerical optimization frameworks to find the locally-optimal parameters.

\section{Numerical Results}
In this section, we evaluate all the preemption policies considered in this work. To illustrate the differences between them, we use the Weibull distribution where $\mathds{P}(Y>k)=\alpha^{k^\beta}$ and set $M=8$ to reduce the parameter space.  

\begin{figure}
    \centering
    \includegraphics[width=0.68\linewidth]{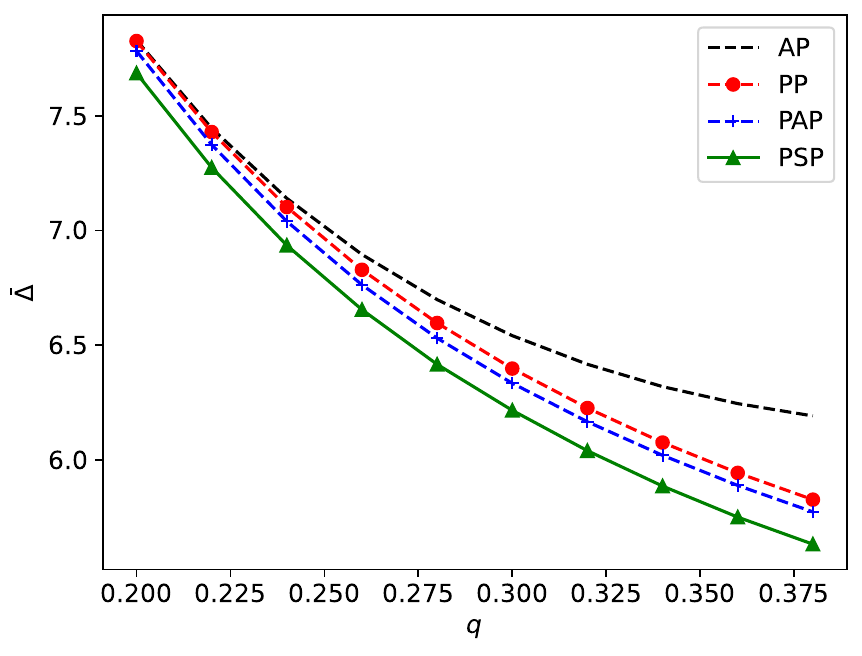}
    \caption{Variation of $\bar \Delta$ with the arrival probability $q$ for $\alpha=0.9$ and $\beta=2$.}
    \label{fig:var_q}
\end{figure}

We evaluate the three policies described in Section \ref{sec:prem_pol} along with the \emph{always preempt} policy (AP). In the first experiment, we set $\alpha=0.9$, $\beta=2$ and vary the arrival probability $q$. As seen in Fig.~\ref{fig:var_q}, the AP policy exhibits the worse performance among the four policies. Moreover, despite the simplicity of the threshold-based preemption policies, they substantially outperform the PP policy by a good margin. Additionally, we see that PSP policy outperforms the PAP policy by a significant margin highlighting the importance of utilizing both the AoP and AoI for the design of the preemption policies.

Next, we highlight some of few key structural features observed in these policies as we vary $q$ for this particular distribution. First, we see that as $q$ increases, the optimal preemption probability of the PP policy decreases. Additionally, we observe that the PAP policy would not preempt for low packet ages and would always preempt for higher packet ages. Further, the threshold $\delta'$ in which the PAP policy shifts between these two strategies increases as we increased $q$. In the PSP policy, we observe that the policy favors preemption at lower AoI values, whereas for higher AoI values, it favors preemption only at higher packet ages. As we increase $q$, the $\Gamma$ at which it shifts between these two strategies decreases. These structural characteristics are very much dependent on the distribution of $Y$. Nevertheless, they highlight the importance of utilizing both the AoP and AoI for the design of the preemption policy.

\begin{figure}
    \centering
    \includegraphics[width=0.68\linewidth]{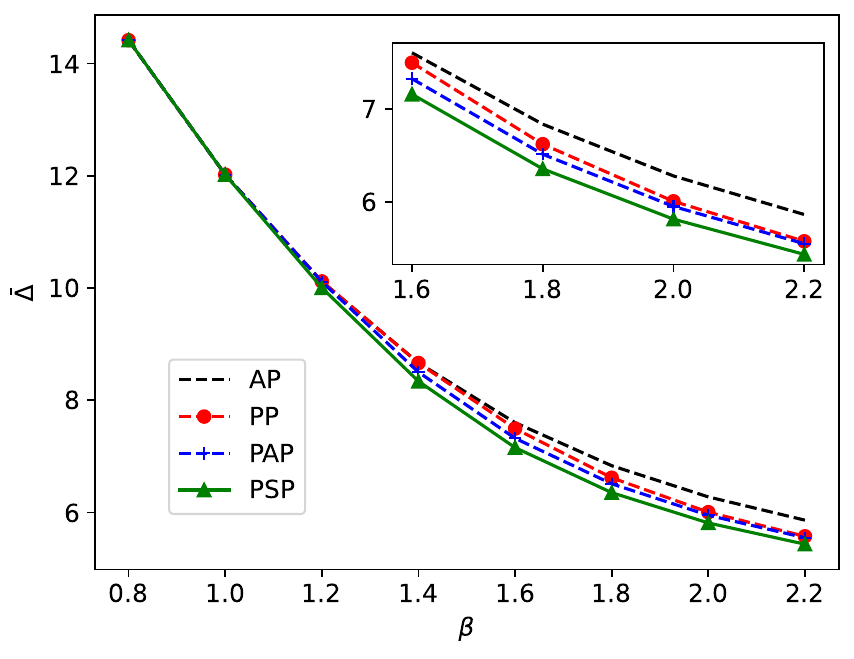}
    \caption{Variation of $\bar \Delta$ with $\beta$ of the Weibull distribution with $\alpha=0.9$ and $q=0.35$.}
    \label{fig:var_beta}
\end{figure}

Finally, we compare how the preemption policies behave as we vary the $\beta$ of the Weibull distribution. It is important to note that for, $\beta<1$, the $y_n$ values are decreasing with $n$ and are increasing if $\beta>1$. Moreover, $\beta=1$ corresponds to the geometric distribution. Fig.~\ref{fig:var_beta} illustrates the variation of $\bar{\Delta}$ as we vary $\beta$ for a fixed $q$. As seen, when $\beta<1$, all policies are identical to AP policy and as $\beta$ increases they start to diverge from the AP policy. This is a potential indicator that an equivalent condition to Theorem \ref{thrm:always_preempt} can be obtained for $q<1$ when $y_n$s are decreasing.

\section{Conclusion}
In this work, we give an analytical framework for computing AoI of multi-threshold preemption policies. Using our analytical framework, we show the importance of utilizing both AoI and AoP when designing these preemption policies.

\appendices
\section{Proof of Proposition \ref{prop:dist_tau_gamma}} \label{appen:prop_dist_tau_gamma}
Let $X_n$ denote the state of the AMC after $n$ time units, and let $\mathcal{S}$ denote the set of transient states and $\mathcal{A}$ denote the set of absorbing states. Let $\bm 1_{M+1}$ be a column vector of all ones of dimension $M+1$ and $\bm 0_M$ be a column vector of all zeros of dimension $M$. Since the initial AoI is $\gamma$, the transition matrix of our non-homogeneous AMC will shift from $\bm P_i$ to $\bm P_{i+1}$ after $\Gamma^{(i)}_\gamma$ time units has elapsed. Moreover, since our AMC always starts upon the successful reception of a packet, we have that the initial distribution our AMC is $[\bm \alpha_0 ~ \bm{0}_M^T]$. Now, for $n\leq \Gamma^{(1)}_\gamma$, we have,
\begin{align}
     \mathds{P}(\tau_\gamma>n)&=\mathds{P}(X_n\in \mathcal{S}|X_0=idle)\\
     &=\begin{bmatrix}
         \bm\alpha_0& \bm0_M^T
     \end{bmatrix}\bm{P}_1^n\begin{bmatrix}
         \bm 1_{M+1}\\
         \bm 0_M
     \end{bmatrix}\\
     &=\begin{bmatrix}
         \bm\alpha_0& \bm0_M^T
     \end{bmatrix}\begin{bmatrix}
         \bm S_1^n &\sum_{k=0}^{n-1}\bm S_1^k\bm A_1\\
         0&\bm I_M
     \end{bmatrix}
     \begin{bmatrix}
         \bm 1_{M+1}\\
         \bm 0_M
     \end{bmatrix}\!\!\\
     &=\bm\alpha_0 \bm S_1^n\bm1.
\end{align}
Then, for $\Gamma^{(1)}_\gamma<n\leq \Gamma^{(2)}_\gamma $, we have,
\begin{align}
    \mathds{P}(\tau_\gamma>n)&=\mathds{P}(X_{\Gamma^{(1)}_\gamma}\in \mathcal{S})\mathds{P}(X_n\in \mathcal{S}|X_{\Gamma^{(1)}_\gamma}\in S)\\
    &=\begin{bmatrix}
         \bm\alpha_0& \bm0_M^T
     \end{bmatrix}\bm P_1^{\Gamma^{(1)}_\gamma} \bm P_2^{n-\Gamma^{(1)}_\gamma}
     \begin{bmatrix}
         \bm 1_{M+1}\\
         \bm 0_M
     \end{bmatrix}\\
     &=\bm\alpha_0\bm S_1^{\Gamma^{(1)}_\gamma} \bm S_2^{n-\Gamma^{(1)}_\gamma}\bm{1}\\
     &=\bm\alpha_0\bm S_1^{\zeta^{(1)}_\gamma} \bm S_2^{n-\Gamma^{(1)}_\gamma}\bm{1}.
\end{align}
Extending this argument for all $i$ yields the desired result.

\section{Proof of Proposition \ref{prop:pi}}\label{appen:prop_pi}
Let $W_n$ be the AoI after the $n$th successful reception of a packet. Note that, $W_n$ forms a discrete-time Markov chain with $M$ states and is irreducible if $y_0>0$. If $y_0=0$, we will have an irreducible chain with fewer states, which can be modeled accordingly. Either way, $W_n$ has a unique stationary distribution which will be the distribution of $\gamma$. Let $\bm \pi$ be this stationary distribution. Next, we find the transition probabilities of $W_n$ as follows,
\begin{align}
     \mathds{P}(W_{n+1}&=j|W_n=i)\nonumber\\
     =&\mathds{P}(X_{\infty}=j_A|\gamma=i,X_0=idle)\\
     =&\sum_{l=1}^{N}\mathds{P}(X_{\Gamma_i^{(l-1)}}\in\mathcal{S},X_{\Gamma_i^{(l)}}=j_A|X_0=idle)\\
     =&\bm\alpha_{0}\sum_{l=1}^{N-1}\left[\left(\prod_{k=0}^{l-1}\bm S_k^{\zeta_i^{(k)}} \right)\sum_{k=0}^{\zeta_i^{(l)}-1}\bm S_l^k\bm A_l\right]\bm \beta_j
     \nonumber\\
     &+\bm\alpha_{0}\left(\prod_{k=0}^{N-1}\bm S_k^{\zeta_i^{(k)}} \right)(\bm I- \bm S_N)^{-1}\bm A_N\bm \beta_j\\
     =&B_{ij}.
\end{align}
Therefore, $\bm \pi$ satisfies, $\bm \pi=\bm \pi B$.

\section{Proof of Theorem \ref{thrm:deter}}\label{appen:thrm_deter}
Let $\Gamma_i=i+1$ for $i\geq 1$. Thus, each $(\Delta_n,\delta_n)$ is associated with a unique $p^{(i)}_k$ preemption probability. Suppose in the optimal policy, $p^{(i)}_k\notin\{0,1\}$. Thus, every time we reach the state $(\Delta_n,\delta_n)=(i,k)$, our AoI curve would branch based on whether we preempted or not. Let $U_1$ denote the average area under the curve starting from state $(i,k)$, until we reach state $(i,k)$ again, if we took the preemption action at state $(i,k)$. Similarly, define $U_2$ to be the area, if we take the action not to preempt. Let $V_1$ and $V_2$ denote the expected cycle lengths for the above two scenarios. Then, we have,
\begin{align}
    \bar\Delta=\frac{p^{(i)}_k U_1+(1-p^{(i)}_k)U_2}{p^{(i)}_k V_1+(1-p^{(i)}_k)V_2}
    \geq\min\left\{\frac{U_1}{V_1},\frac{U_2}{V_2}\right\}.
\end{align}
Since $\frac{V_1}{W_1}$ is the average AoI of a policy which always preempts when $(\Delta_n,\delta_n)=(i,k)$, and $\frac{V_2}{W_2}$ is the average AoI for a policy which never preempts in the same state, we have that one of the above two policies is better than or equal to our optimal policy. Applying this to all $p^{(i)}_k\notin\{0,1\}$, we can contruct an equally optimal or better deterministic policy.

\section{Proof of Theorem \ref{thrm:always_preempt}}\label{appen:thrm_always_preempt}
If $y_0=1$, the result is trivial. Now, consider the case $y_0<1$. From \eqref{eqn:avg_age}, \eqref{eqn:1_moment} and \eqref{eqn:2_moment}, we have that,
\begin{align}
    \bar{\Delta}&\geq\min_\gamma\left\{\gamma+\frac{\e[\tau_\gamma^2]}{2\e[\tau_\gamma]}-\frac{1}{2}\right\}\\
    &=\min_\gamma\left\{\gamma+\frac{\e[\tau_\gamma]+2\sum_{k=1}^\infty k \mathds{P}(\tau_\gamma>k)}{2\e[\tau_\gamma]}-\frac{1}{2}\right\}\\
    &=\min_\gamma\left\{\gamma+\frac{\sum_{k=1}^\infty k \mathds{P}(\tau_\gamma>k)}{1+\sum_{k=1}^\infty  \mathds{P}(\tau_\gamma>k)}\right\}.\label{eqn:delta_bound}
\end{align}

Let $\bar \Delta_\gamma=\gamma+\frac{\sum_{k=1}^\infty k \mathds{P}(\tau_\gamma>k)}{1+\sum_{k=1}^\infty  \mathds{P}(\tau_\gamma>k)}$. Next, for a given $\gamma$, we find the optimal preemption policy that minimizes $\bar \Delta_\gamma$. Let this minimum value be denoted by $\bar \Delta_\gamma^*$. Note that, $\tau_\gamma$ is the time for absorption starting from the $idle$ state. Since we have fixed $\gamma$, our actions, i.e., the preemption probabilities, will depend only on the time elapsed since the AMC began and the current age of the packet. Thus, let $p_{i,j}$ denote the preemption probability given that the time elapsed is $i$ and the packet age is $j$ ($j=0 \equiv idle$). Let $z_{i,j}$ be the probability that the packet age is $j$ given the AMC has not been absorbed after $i$ time units have elapsed. Then, $z_{0,0}=1$. Now, given that the AMC has not been absorbed by time $i$, the probability that it will be absorbed in the next time step is given by $x_i=z_{i,0}p_{i,0}y_0+\sum_{j=1}^Mz_{i,j}(p_{i,j}y_0+\bar p_{i,j}y_j)$. Then, we have the following,
\begin{align}
    \mathds{P}(\tau_\gamma>0)&=1\\
    \mathds{P}(\tau_\gamma>k)&=\mathds{P}(\tau_\gamma>k-1)\mathds{P}(\tau_\gamma>k|\tau_\gamma>k-1)\\
    &=\mathds{P}(\tau_\gamma>k-1)(1-x_{k-1})\\
    &=\prod_{i=0}^{k-1}(1-x_i).
\end{align}

Now, minimizing $\bar\Delta_\gamma$ is equivalent to the following optimization problem,
\begin{mini}
    {p_{i,j}\in[0,1]}{\frac{\sum_{k=1}^{\infty}k\prod_{i=0}^{k-1}(1-x_i)}{1+\sum_{k=1}^{\infty}\prod_{i=0}^{k-1}(1-x_i)}}
    {\label{opt:org_opt}}
    {}.
\end{mini}
Let the minimum of the above optimization problem be denoted by $J(\gamma)$. In the above optimization problem, the $x_i$s are correlated through the actions $p_{i,j}$ that we take. Further, note that, since $y_j\leq y_0$, we have that $x_i\leq  y_0$ for any feasible preemption policy.  Hence, if we select the particular $x_i$s independently subject to the constraint $x_i\leq y_0$, we can obtain a lower bound $L(\gamma)$ such that $J(\gamma)\geq L(\gamma)$, where $L(\gamma)$ is the solution to the following optimization problem,
\begin{mini}
    {x_i\geq0}{\frac{\sum_{k=1}^{\infty}k\prod_{i=0}^{k-1}(1-x_i)}{1+\sum_{k=1}^{\infty}\prod_{i=0}^{k-1}(1-x_i)}}
    {\label{opt:LB_opt}}
    {}
    \addConstraint{x_i}{\leq  y_0}.
\end{mini}
To obtain the lower bound, we first select an $x_i$ (say $x_j$) and find the optimizing $x_j$ for a given set of $x_i$s for $i\neq j$. To do this, we rearrange the above objective function as follows, 
\begin{align}
    &\!\!\!\!\!\frac{\sum_{k=1}^{\infty}k\prod_{i=0}^{k-1}(1-x_i)}{1+\sum_{k=1}^{\infty}\prod_{i=0}^{k-1}(1-x_i)}\nonumber\\
    &\!\!\!\!\!=j+1-\frac{\left(j+1+\sum_{k=1}^{j}(j+1-k)\prod_{i=0}^{k-1}(1-x_i)\right)}{1+\sum_{k=1}^{\infty}\prod_{i=0}^{k-1}(1-x_i)}\nonumber\\
    &+\frac{\sum_{k=j+1}^{\infty}(k-j-1)\prod_{i=0}^{k-1}(1-x_i)}{1+\sum_{k=1}^{\infty}\prod_{i=0}^{k-1}(1-x_i)}\\
    &\!\!\!\!\!=j+1-\frac{\left(j+1+\sum_{k=1}^{j}(j+1-k)\prod_{i=0}^{k-1}(1-x_i)\right)}{1+\sum_{k=1}^{\infty}\prod_{i=0}^{k-1}(1-x_i)}\nonumber\\
    &+\frac{1-x_{j+1}+\sum_{k=j+3}^{\infty}(k-j-1)\prod_{i=j+1}^{k-1}(1-x_i)}{\frac{1+\sum_{k=1}^{j}\prod_{i=0}^{k-1}(1-x_i)}{\prod_{i=0}^{j}(1-x_i)}+1+\sum_{k=j+2}^{\infty}\prod_{i=j+1}^{k-1}(1-x_i)}\!\!
\end{align}
with the convention that $\sum_{k=i}^l$ operator results in zero for $l<i$. Next, we note that, $x_j$ is present only in the denominator of the second fraction through the term $(1-x_j)$ and in the denominator of the third fraction as $\frac{1}{(1-x_j)}$. To minimize the above expression, we need to maximize the magnitude of the negative fraction and minimize the magnitude of the positive fraction. This can be jointly achieved at the same time by setting $x_j=y_0$. Therefore, for a fixed set of $x_i$s ($i\neq j)$, we can minimize the above expression by setting $x_j=y_0$ and this choice is independent of the rest of the values we select for $x_i$s. Therefore, $x_i=y_0~\forall i$ achieves the minimum $L(\gamma)$.

Next, we note that if we set $p_{i,j}=1$ in \eqref{opt:org_opt}, we get $x_i= y_0$. Hence, the choice $p_{i,j}=1$ achieves the minimum of \eqref{opt:org_opt}. Therefore, $J(\gamma)$ is minimized by an always preempt (AP) policy. This yields that $\bar \Delta_\gamma^*=\gamma+\frac{1-y_0}{y_0}$. Therefore, from \eqref{eqn:delta_bound}, we have that $\bar \Delta\geq \bar \Delta_1^*$. Finally, we observe that under an AP policy $\pi_1=1$. Therefore, $\bar \Delta=\Delta_1^*$, under the AP policy. This proves the desired result.

\bibliographystyle{unsrt}
\bibliography{refs}

@ARTICLE{yates2020age,
    author      ={R. D. Yates and Y. Sun and D. R. Brown and S. K. Kaul and E. Modiano and S. Ulukus},
    title       ={Age of Information: An Introduction and Survey},
    journal     ={IEEE Journal on Selected Areas in Communication}, 
    year        ={2021},
    month       ={May},
    volume      ={39},
    number      ={5},
    pages       ={1183-1210}
}

@INPROCEEDINGS{subhankar_preempt,
	author      ={S. Banerjee and S. Ulukus},
	booktitle   ={IEEE ISIT},
	title       ={When to Preempt in a Status Update System?},
	year        ={2024},
	month      	={July}
}

@ARTICLE{elif_heavytail, 
      title={Taming the Heavy Tail: Age-Optimal Preemption}, 
      author={A. Li and Y. Ince and E. Uysal},
      note={Available online at arXiv:2601.16624}}

@INPROCEEDINGS{prob_preempt,
	author      ={Moltafet, M. and Sadjadpour, H. R. and Rezki, Z. and Codreanu, M. and Yates, R. D.},
	booktitle   ={IEEE ISIT},
	title       ={AoI in {M/G/1/1} Queues with Probabilistic Preemption},
	year        ={2025},
	month      	={June}
}

@ARTICLE{link_capacity_preempt,
    author      ={Wang, B. and Feng, S. and Yang, J.},
    title       ={When to preempt? Age of information minimization under link capacity constraint},
    journal     ={Journal of Communications and Networks}, 
    year        ={2019},
    month       ={June},
    volume      ={21},
    number      ={3},
    pages       ={220-232}
}

@ARTICLE{delayed_premption,
    author      ={Asvadi, S. and Ashtiani, F.},
    title       ={Delayed Preemption: A New Policy for Balancing the Age of Information in Prioritized Networks},
    journal     ={IEEE Transactions on Communications}, 
    year        ={2025},
    month       ={October},
    volume      ={73},
    number      ={12},
    pages       ={14639-14652}
}

@INPROCEEDINGS{timely_computing_preempt,
	author      ={Arafa, A. and Yates, R. D. and Poor, H. V.},
	booktitle   ={Allerton Conference},
	title       ={Timely Cloud Computing: Preemption and Waiting},
	year        ={2019},
	month      	={September}
}

@INPROCEEDINGS{status_update_queues,
	author      ={Kaul, S. K. and Yates, R. D. and Gruteser, M.},
	booktitle   ={CISS},
	title       ={Status updates through queues},
	year        ={2012},
	month      	={March}
}
\end{document}